\documentclass[sigconf,screen=true,bookmarks=true]{acmart}

\AtBeginDocument{%
  \providecommand\biburl[1]{}%
  \renewcommand\biburl[1]{}%
  \providecommand\url[1]{}%
  \renewcommand\url[1]{}%
}
\usepackage{graphicx}
\usepackage{amsmath}
\usepackage{mathtools}
\usepackage{comment}
\usepackage[subrefformat=parens,labelformat=parens]{subfig}
\usepackage{bm}
\usepackage{multirow}
\usepackage{threeparttable,booktabs}
\usepackage{tikz}
\usepackage{balance}
\usepackage{courier}                                       
\usepackage{cleveref}                                      
\usepackage[mathcal]{eucal}
\usepackage[noend]{algpseudocode}
\algrenewcommand\textproc{\texttt}
\makeatletter\let\float@addtolists\relax\makeatother
\usepackage[ruled,linesnumbered]{algorithm2e}
\usepackage{filecontents}                                  
\usepackage{pgfplots}
\usepackage{pgfplotstable}
\pgfplotsset{compat=newest}
\usepackage{siunitx}
\theoremstyle{plain}

\theoremstyle{definition}

\newcommand{\tabRef}[1]{Table~\ref{#1}}

\usepackage[skip=1pt]{caption}            
\graphicspath{{./figs/}}
\usepackage{algorithmicx}
\usepackage{algpseudocode}
\usepackage{threeparttable}
\usepackage[most]{tcolorbox}

\newcommand\blfootnote[1]{%
  \begingroup
  \renewcommand\thefootnote{}\footnote{#1}%
  \addtocounter{footnote}{-1}%
  \endgroup
}

\acmDOI{10.1109/DAC63849.2025.11133257}
\acmISBN{}
\begin{document}


\title[PANDA: An LLM-Enhanced Performance-Driven Analog Design Framework]{
Special Research Session: PANDA: An LLM-Enhanced Performance-Driven Analog Design Framework Bridging Design Intent and Layout Generation 
}

\author[]{
  Haoyi Zhang$^{1}$, Weijian Fan$^{1}$, Xiaohan Gao$^{1}$, Bingyang Liu$^{1}$, Runsheng Wang$^{1,2,3}$, Yibo Lin$^{1,2,3*}$
}
\affiliation{
\institution{
\large{$^{1}$School of Integrated Circuits, Peking University, Beijing, China} \\
\large{$^{2}$Beijing Advanced Innovation Center for Integrated Circuits, Beijing, China} \\
\large{$^{3}$Institute of Electronic Design Automation, Peking University, Wuxi, China}}
\country{}
}
 \email{{hy.zhang@pku.edu.cn, yibolin@pku.edu.cn}}

\begin{abstract}

\begin{sloppypar}
Traditional design of analog circuits heavily relies on manual interventions across topology, sizing, and layout, with prior automation addressing stages in isolation. In this work, we propose PANDA, an LLM‑enhanced framework that bridges high‑level design intent to final layout by actively managing cross‑stage dependencies through guided topology synthesis, substructure‑aware sizing, and constraint‑driven layout generation. This shifts automation from algorithm‑centric execution to intent‑centric co‑design, reducing turnaround time from days or weeks to hours while improving design performance. PANDA is released at \textcolor{blue}{\texttt{https://github.com/PKU-IDEA/PANDA}}.
\end{sloppypar}

\end{abstract}

\maketitle

\blfootnote{$^*$Corresponding author. DOI: 10.1109/DAC63849.2025.11133257}
\section{Introduction}
\label{sec:Introduction}

\begin{sloppypar}
Analog and mixed-signal (AMS) design automation is increasingly critical as circuit complexity grows, yet despite decades of analog CAD research, practical end‑to‑end automation remains limited. As illustrated in the traditional flow of Figure 1 (top), topology synthesis, sizing, and layout are handled in a loosely coupled manner, leaving three persistent gaps. First, conventional algorithms operate without understanding design intent, performing inefficient, directionless exploration that often misses optimal regions~\cite{ILAC, KOAN, crossley_bag_2013, dhar_align_2020, MAGICAL, chen_magical_2021, martins_analog_2023,11126279,zhang_sageroute_nodate,zhangSAGERoute2.0SynergisticAnalog}. Second, schematic and layout stages are isolated: sizing is optimized without layout awareness, while layout tools lack structured links to circuit‑level decisions, breaking optimization loops~\cite{bhattacharya_template-driven_nodate, lourenco_laygen_2006, lourenco_layout-aware_2015, ou16tcad_lde, zhang_parasitic-aware_2008, liu_parasitic-aware_2021, xu_performance-driven_2024, torabi_electromigration-_2018}. Third, existing flows cannot autonomously iterate based on post‑layout simulation results; manual intervention is required to extract parasitics, re‑evaluate performance, and adjust parameters, preventing true performance‑driven refinement.
\end{sloppypar}

\begin{figure}[htbp]
  \centering
  \includegraphics[width=0.65\linewidth]{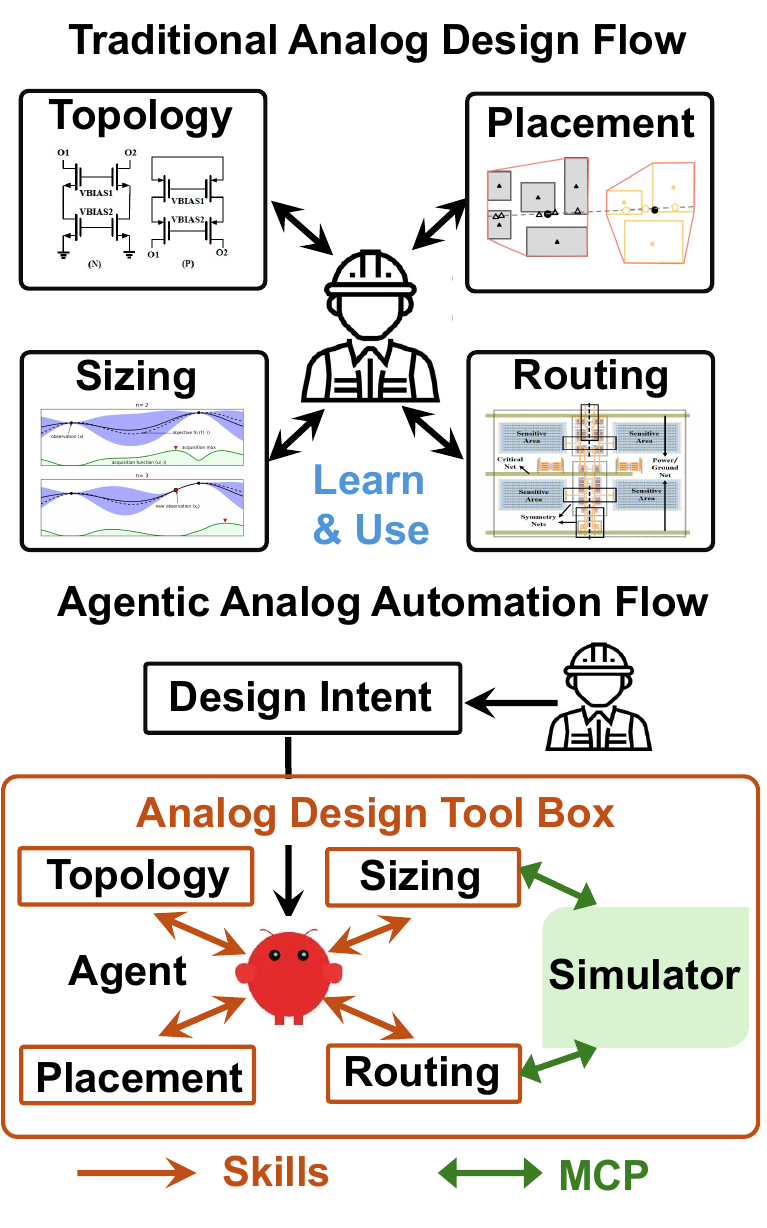}
  \caption{Comparison between traditional automated analog design flow and the agentic flow in PANDA.}
  \Description{A two-part comparison between a fragmented traditional analog automation flow and PANDA's agentic flow, highlighting tighter coordination across topology, sizing, placement, routing, and post-layout feedback in PANDA.}
  \label{fig:intro-overview}
  \vspace{-0.5cm}
\end{figure}

\begin{sloppypar}
Prior research has made important advances in individual stages, including topology synthesis~\cite{meissnerFEATSFrameworkExplorative2015, zhaoAutomatedTopologySynthesis2020, luTopologyOptimizationOperational2022}, transistor sizing~\cite{sizing2, sizing3, sizing4, sizing1, settaluriAutomatedDesignAnalog2022}, placement~\cite{11126279,balasa02iccad_sym, ma11tcad_constraints, xu19ispd_place, zhu20iccad_ssf, dhar22date_charge}, and routing~\cite{chen_toward_2020_iccad, zhang_sageroute_nodate, zhangSAGERoute2.0SynergisticAnalog, xu_performance-driven_nodate}. Yet these efforts remain fragmented, and the integration of high‑level design intent with physical implementation is still missing. Recent large language models (LLMs) offer new opportunities to bridge these gaps~\cite{chen24artisan,dubey2024llama3herdmodels,heChatEDALargeLanguage2024,laiAnalogCoderAnalogCircuit2024,10841395,liuChipNeMoDomainAdaptedLLMs2024,madaan2023selfrefineiterativerefinementselffeedback,touvron2023llama2openfoundation,wei2023chainofthoughtpromptingelicitsreasoning,yinADOLLMAnalogDesign2024}
, but intent understanding alone is insufficient without a framework that preserves cross‑stage dependencies and grounds generated plans in deterministic tool executions.
\end{sloppypar}

\begin{sloppypar}

PANDA addresses these challenges by employing an LLM as a design coordinator that interprets intent, manages cross‑stage dependencies, and drives iterative improvement. As shown in the proposed agentic flow of Figure 1 (bottom), PANDA unifies topology synthesis, analog sizing, placement, and routing into a coherent pipeline. The LLM translates specifications into structured constraints so that sizing follows topology, placement respects device‑level requirements, and routing receives consistent interface definitions. After layout, PANDA automatically extracts parasitics and feeds post‑layout performance back to the LLM, enabling it to select the best implementation and iteratively refine the design. By bridging schematic and layout with a semantically aware coordinator, PANDA shifts analog automation from isolated, algorithm‑centric execution to an intent‑centric, self‑improving flow.
\end{sloppypar}


In summary, this work makes four key contributions:
\begin{itemize}
  \begin{sloppypar}
    \item \textbf{Agent Flow}: Agentic analog design flow that connects high‑level design intent to final physical implementation through four coordinated stages, preserving the dependency chain from circuit semantics to executable layout actions.
    \item \textbf{Topology Synthesis}: LLM‑guided topology synthesis that translates compact design intent and specifications into explicit structural artifacts, providing a stable schematic representation for downstream optimization.
    \item \textbf{Analog Sizing}: Substructure‑aware analog sizing using a multi‑stage hierarchical Bayesian optimization engine that resolves device parameters from the generated topology and emits transistor‑level artifacts directly consumable by layout.
    \item \textbf{Layout Generation}: Constraint‑driven layout generation that combines placement and routing with schematic‑layout synchronization, converting structural and sizing information into geometry‑aware constraints and executable back‑end actions.
  \end{sloppypar}
\end{itemize}
Based on these contributions, PANDA demonstrates a unified path toward practical analog design automation, reducing turnaround time from days or weeks to hours while improving design performance. By bridging design intent and physical realization with an LLM‑coordinated flow, PANDA represents a significant step toward closing the loop in analog design automation.


\section{Background and Motivation}
\label{sec:background}

\begin{sloppypar}
Analog automation remains substantially less mature than digital design automation because the information required for design closure is distributed across multiple heterogeneous stages. While digital flows benefit from standardized abstractions and clean stage interfaces, analog design must continuously reconcile circuit intent, device-level performance, matching constraints, parasitic effects, and layout regularity.
\end{sloppypar}

\subsection{Breakpoints in Analog Design Closure}
\begin{sloppypar}
Conventional analog flows contain three persistent breakpoints~\cite{dhar_align_2020, chen_magical_2021, xu_performance-driven_2024, lin_parasitic-aware_2017}. First, the transition from user-level design intent to a feasible circuit topology is still highly manual and experience-driven. Second, transistor sizing is often optimized in isolation, making it difficult to transfer performance-critical information to later physical stages. Third, placement and routing require explicit geometric and electrical constraints, such as symmetry, common-centroid structures, and consistent pin definitions, which are rarely generated in a unified way from earlier design decisions. As a result, the flow is often broken into loosely coupled steps that require repeated manual repair.
\end{sloppypar}



\subsection{From Agent Reasoning to Executable Flows}
\begin{sloppypar}
Recent agentic systems suggest that LLM-based workflows become substantially more reliable when reasoning is decomposed into reusable task units and connected to explicit tool interfaces~\cite{heChatEDALargeLanguage2024, baekResearchAgentIterativeResearch2024}. In this context, a \textit{Skill} can be viewed as a stage-specialized capability that encapsulates prompting, context construction, and post-processing for a particular design task, while an MCP-style interface acts as a normalized protocol for invoking heterogeneous external tools and services. For analog automation, these abstractions are attractive because they separate semantic planning from physical execution without breaking the underlying stage dependencies.
\end{sloppypar}

\begin{sloppypar}
PANDA adopts this perspective in a lightweight way. Skills provide structured reasoning entry points for topology, sizing, placement, and routing, whereas MCP-style invocations provide stable execution channels to the back-end engines. We introduce these concepts here as background because they motivate how PANDA connects LLM guidance with deterministic analog design closure; their concrete role in the framework is detailed in Section~\ref{sec:Algorithm}.
\end{sloppypar}

\section{PANDA Methodology}
\label{sec:Algorithm}

\begin{figure*}[t]
  \centering
  \includegraphics[width=0.98\textwidth]{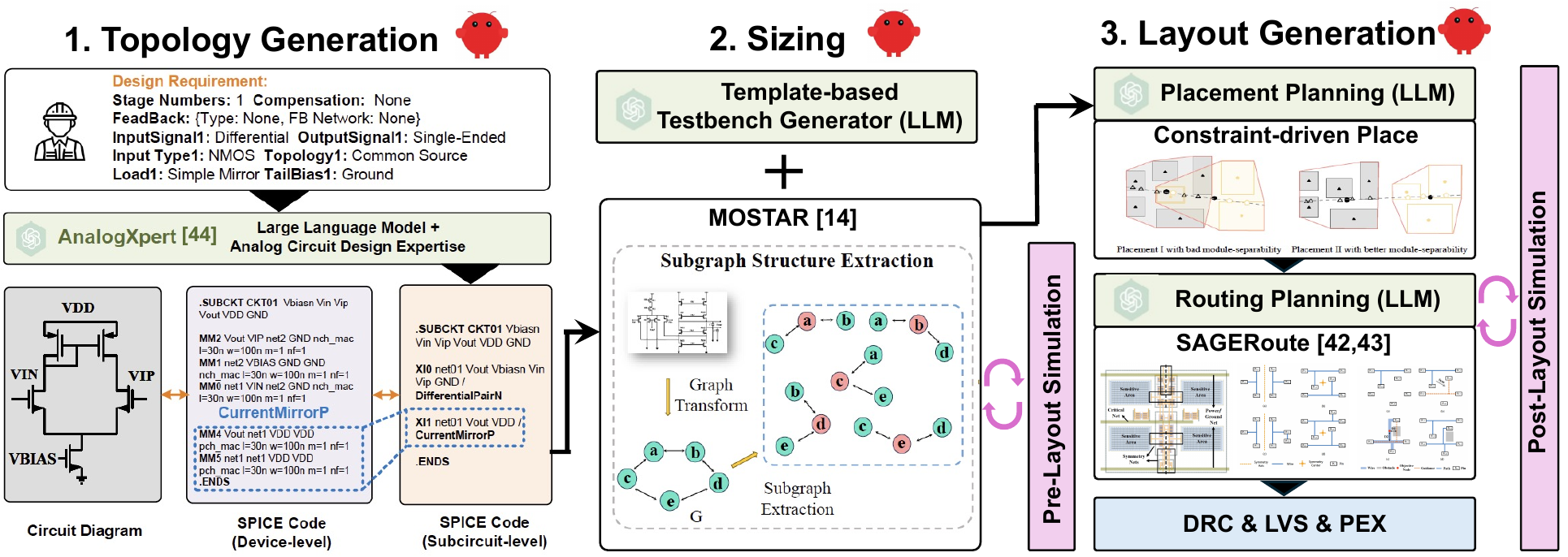}
  \caption{Overview of PANDA methodology.}
  \Description{A full-width workflow diagram of PANDA showing inputs design intent and design specifications, followed by topology synthesis, analog sizing, placement, routing, and post-layout feedback among reusable Skills and back-end engines.}
  \label{fig:method-overview}
\end{figure*}

\begin{sloppypar}
PANDA is designed as an integrated framework that connects high-level design intent to final analog layout through four tightly coupled stages: topology synthesis, analog sizing, placement, and routing. Across these stages, the LLM serves as a planner and constraint generator, while the back-end engines execute physically grounded design actions. The methodology is therefore centered on progressive context refinement: each stage transforms design intent into more concrete structural, parametric, and geometric representations until executable layout actions are obtained.
\end{sloppypar}

\subsection{Framework Overview}
\begin{sloppypar}
Figure~\ref{fig:method-overview} presents the overall PANDA workflow. The framework takes compact user inputs (\texttt{design\_intent} and \texttt{design\_spec}) and drives a four-stage refinement chain: topology synthesis, analog sizing, constraint-driven placement, and routing with schematic-layout synchronization. Rather than treating these stages as isolated tools, PANDA organizes them as a dependency-preserving pipeline in which each stage consumes structured outputs from the previous one and emits execution-ready artifacts for the next.
\end{sloppypar}

\begin{sloppypar}
At the control level, PANDA separates semantic planning from deterministic execution. The LLM is responsible for interpreting intent, selecting stage-specific actions, and generating machine-readable constraints, while specialized back-end engines perform simulation, optimization, placement, and routing. This division ensures that high-level reasoning remains flexible, but physically grounded decisions are always produced by domain tools with explicit interfaces.
\end{sloppypar}

\begin{sloppypar}
To make this interaction reproducible, PANDA implements the workflow as reusable \textit{Skills} connected through explicit handoff artifacts. As summarized in \tabRef{tab:skill_abstraction}, these Skills cover generation, decision, planning, and execution tasks across all stages. Together, they form a closed-loop automation process that preserves cross-stage consistency and supports iterative improvement from post-layout feedback.
\end{sloppypar}

\begin{table}[htbp]
    \centering
    \caption{Abstracted task Skills in PANDA.}
    \label{tab:skill_abstraction}
    \small
    \setlength{\tabcolsep}{3pt}
    \renewcommand{\arraystretch}{1}
    \begin{tabular}{>{\centering\arraybackslash}m{0.27\linewidth}>{\centering\arraybackslash}m{0.44\linewidth}>{\centering\arraybackslash}m{0.19\linewidth}}
        \toprule
        Task & Role & Main output \\
        \midrule
        \texttt{topology\_} \newline \texttt{generation} & Generates the initial circuit topology from intent and specs. & \texttt{topology.json} \\
        \addlinespace[1pt]
        \texttt{post\_topology\_} \newline \texttt{decision} & Dispatches downstream tasks and derives stage-specific hints. & planning JSONs \\
        \addlinespace[1pt]
        \texttt{sizing} & Resolves device parameters for target performance. & \texttt{sizing.json} \\
        \addlinespace[1pt]
        \texttt{placement\_pr\_} \newline \texttt{planning} & Builds placement constraints such as grouping and symmetry. & \texttt{placement\_} \newline \texttt{pr.json} \\
        \addlinespace[1pt]
        \texttt{placement} & Executes physical placement using planned constraints. & \texttt{pl.json} \\
        \addlinespace[1pt]
        \texttt{routing\_pr\_} \newline \texttt{planning} & Prepares routing directives and interface constraints. &\texttt{routing\_} \newline \texttt{pr.json}  \\
        \addlinespace[1pt]
        \texttt{routing} & Executes detailed routing and final layout connectivity. & \texttt{rt.json} \\
        \bottomrule
    \end{tabular}
\end{table}

\subsection{Topology Synthesis}
\begin{sloppypar}
In the first stage, PANDA invokes \textit{AnalogXpert} to synthesize circuit topology as subcircuit-level SPICE generation from structured requirements (function, performance targets, and interface constraints)~\cite{analogxpert}. The agent narrows the search space with an extensible subcircuit library, constructs connectivity with iterative proofreading, and outputs a practical \texttt{.SUBCKT} representation. PANDA then parses this result into structured artifacts (netlist, device relations, and terminal mappings) and stores them in \texttt{topology.json}; if generation is invalid, it falls back to the closest valid subcircuit combination and re-runs refinement to recover a consistent topology for downstream stages.
\end{sloppypar}

\subsection{Analog Sizing}
\begin{sloppypar}
With topology fixed, PANDA performs transistor-level sizing as constrained optimization over parameters such as width, length, finger number, and multipliers. The framework uses \textit{MOSTAR}, a multi-stage hierarchical Bayesian optimizer that combines substructure-aware modeling (via L2G-GNN), symmetric constraints, additive Gaussian processes, and stage-adaptive acquisition to handle high-dimensional black-box objectives efficiently~\cite{fan_mostar_2025}. Optimization iteratively evaluates candidate designs and updates the surrogate model, and the final execution-ready parameter set is saved as \texttt{sizing.json} for subsequent physical layout stages.
\end{sloppypar}

\subsection{Constraint-Driven Placement}
\begin{sloppypar}
In placement, PANDA translates schematic structure and layout intent into geometry using a constraint-driven placement framework that co-optimizes area/wirelength proxies, symmetry and collinearity constraints, and signal/power-flow patterns. It then applies legalization to remove overlaps and produce a compact routable solution, and stores the final placement artifacts (including \texttt{pl.json}) for consistent downstream routing and quality evaluation.
\end{sloppypar}

\subsection{Routing and Layout Synchronization}
\begin{sloppypar}
In the final stage, PANDA aggregates topology, sizing, and placement outputs to generate routing directives and executes detailed routing with \texttt{SAGERoute}/\texttt{SAGERoute2.0} under unified design-rule and interface constraints~\cite{zhang_sageroute_nodate, zhangSAGERoute2.0SynergisticAnalog}. To maintain schematic-layout consistency, it synchronizes the LVS pin list in \texttt{template\_json/allconfig.json} from topology-derived interfaces, iteratively updates routing configurations, and runs \textit{PEX} after each attempt to obtain parasitic-aware performance/yield feedback; these metrics are then fed back to the LLM to select and refine the best feasible routed solution~\cite{10841395}.
\end{sloppypar}

\section{Experimental Results}
\label{sec:Results}

\begin{sloppypar}
We evaluate PANDA from two perspectives: execution transparency and design quality. In addition to reporting final layout and circuit performance, we show representative intermediate prompts and structured Skill outputs that reveal how the flow progresses from user intent to executable layout actions. We demonstrate PANDA on two cases, \textit{OTA} and \textit{COMP}. Due to page limits, we present the OTA flow in detail and summarize COMP with key metrics.
\end{sloppypar}

\begin{sloppypar}
For OTA, we use the design request ``Design a three-stage OTA with differential input and single output, targeting high gain and robust phase margin for sensor readout,'' with \texttt{design\_spec} \texttt{circuit\_type=ota}, \texttt{target\_gain\_db=75}, and \texttt{target\_ugbw\_mhz=8}. For COMP, we evaluate a StrongARM comparator case and summarize the post-layout power and delay results reported in \tabRef{tab:perf_compare}.
\end{sloppypar}

\subsection{Execution Flow}
\begin{sloppypar}
PANDA executes the design task as a staged pipeline from \texttt{design\_intent} and \texttt{design\_spec} to final layout. The flow first generates a topology artifact, then derives sizing parameters, placement constraints, and routing directives, with each stage emitting an explicit machine-readable output for the next one. These intermediates are stored in attempt-specific files such as \texttt{topology.json}, \texttt{sizing.json}, \texttt{pl.json}, and \texttt{rt.json}, which makes the run traceable and easy to inspect.
\end{sloppypar}

\begin{sloppypar}
To preserve cross-stage consistency, PANDA synchronizes downstream inputs directly from upstream artifacts instead of reconstructing them from free-form text. For example, routing configurations inherit interface information such as \texttt{lvs\_pins} from topology outputs or parsed \texttt{.SUBCKT} definitions, while placement-derived symmetry information is persisted as structured constraints for later stages. This organization keeps the execution chain reproducible and maintains alignment between schematic intent and physical implementation.
\end{sloppypar}

\begin{sloppypar}
In our experiments, the end-to-end runtime for a full PANDA run is at the \textit{several-hours} level, whereas comparable manual analog layout iterations typically require \textit{several days}. This gap highlights the practical turnaround advantage of the staged and observable PANDA flow.
\end{sloppypar}

\subsection{Prompt and Skill Examples}
\begin{sloppypar}
Beyond final artifacts, PANDA also records representative prompt-level intermediates that show how design intent is progressively translated into structured downstream actions. Rather than presenting full prompt transcripts, we highlight three concise examples that correspond to different abstraction levels in the flow.
\end{sloppypar}

\begin{tcolorbox}[colback=gray!5!white, colframe=gray!75!black, title=Example Prompt and Skill Interactions in PANDA, fonttitle=\bfseries, boxrule=0.75pt, arc=2mm, left=3pt, right=3pt, top=3pt, bottom=3pt]
    \footnotesize
    \textbf{Design Intent:} \\
    The flow begins from a compact user-level request: design a three-stage OTA with differential input and single output, targeting high gain and robust phase margin for sensor readout. The corresponding structured context in \texttt{flow\_request.json} includes \texttt{target\_gain\_db=75} and \texttt{target\_ugbw\_mhz=8}. PANDA interprets this request as a flow-level objective and initializes the task scope, context-loading policy, and retry budget.

    \par\smallskip
    \textbf{Post Topology Decision:} \\
    After topology generation, PANDA invokes the post-topology decision Skill to convert the topology summary into downstream actions. For the generated three-stage OTA, the Skill schedules the next agents for sizing, placement, and routing; prepares files such as \texttt{sizing\_spec.json}, \texttt{placement\_constraints.json}, and \texttt{routing\_rules.json}; and derives guidance including symmetry pair \texttt{[XI1, XI2]}, critical devices \texttt{[XI7, XI8]}, and sensitive nets \texttt{[OUT1, OUT2, OUT3]}.

    \par\smallskip
    \textbf{Placement Planning:} \\
    At the placement stage, the placement-planning Skill focuses on analog-layout-specific constraints including grouping, symmetry, matching, and whitespace reservation. The planning prompt explicitly requires that all PMOS devices be grouped, while NMOS devices may remain ungrouped. The resulting \texttt{placement\_pr.json} records clustered placement objects, symmetric pairs such as \texttt{[MM25, MM26]} and \texttt{[MM39, MM40]}, whitespace settings, and an automatic consistency check that reports full PMOS coverage.
   
\end{tcolorbox}

\begin{sloppypar}
These examples show that PANDA does not rely on opaque prompt chaining. Instead, each prompt is tied to a stage-specific Skill and materialized as a concise structured object that can be executed, checked, and propagated to later stages.
\end{sloppypar}

\subsection{Pre-Layout and Post-Layout Performance}

\begin{sloppypar}
    To evaluate design quality, \tabRef{tab:perf_compare} reports pre- and post-layout results for both OTA and COMP. For OTA, the post-layout results (via PEX) retain high gain (76.11 dB) and robust phase margin (65.3$^\circ$), which are consistent with the design intent of high-gain and stable sensor-readout operation. The UGB decreases from 10.84 MHz to 4.192 MHz after layout parasitics are included, so the frequency target is not fully maintained post-layout even though the main amplification behavior is preserved. For COMP, no explicit quantitative performance targets are specified in the design request, so we report power and delay as representative implementation metrics. The post-layout result shows 606.6 nW power and 1.447 ns delay, compared with 469.9 nW and 1.0 ns pre-layout, indicating the expected parasitic-induced overhead while remaining within a practical operating range for a functional StrongARM comparator implementation. Overall, the post-layout results show that PANDA can carry both designs from intent to physically realized implementations with clear and traceable performance tradeoffs.
\end{sloppypar}

\begin{figure}[htbp]
    \centering
    \includegraphics[width=0.6\linewidth]{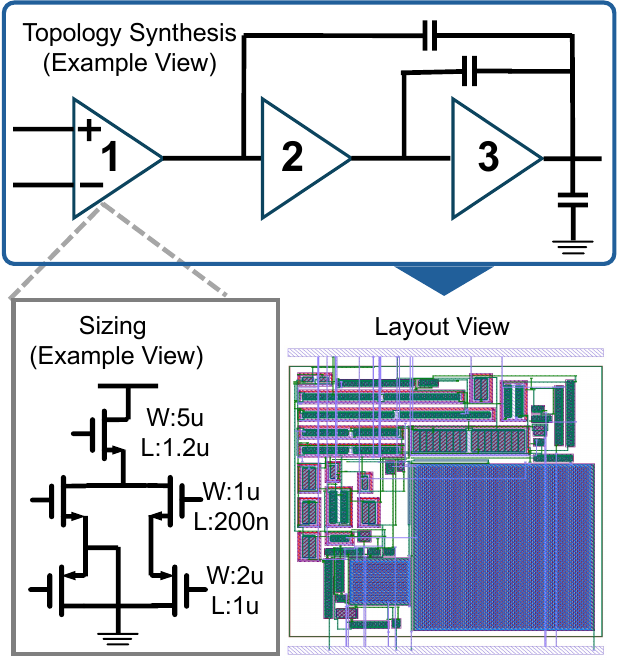}
    \caption{Generated layout result for the OTA case.}
    \Description{The generated OTA layout after placement and routing, illustrating a compact physical implementation produced by PANDA.}
    \label{fig:generated_layout}
\end{figure}

\begin{table}[htbp]
    \centering
    \caption{Pre-layout and post-layout performance comparison for the OTA and COMP case studies.}
    \label{tab:perf_compare}
    \begin{tabular}{llccc}
        \toprule
        Design & Metric & Design intent & Pre-layout & Post-layout \\
        \midrule
        \multirow{4}{*}{OTA} & Gain (dB) & $> 75$ & 79.07 & 76.11 \\
         & PM ($^\circ$) & $> 60$ & 67.47 & 65.3 \\
         & UGB (MHz) & $> 8$ & 10.84 & 4.192 \\
         & Power ($\mu$W) & -- & 411.2 & 435.2 \\
        \midrule
        \multirow{2}{*}{COMP} & Power (nW) & -- & 469.9 & 606.6 \\
         & Delay (ns) & -- & 1.0 & 1.447 \\
        \bottomrule
    \end{tabular}
\end{table}

\subsection{Attempt-Level Analysis}
\begin{sloppypar}
PANDA is designed to be both observable and executable: it logs stage transitions, invocation status, and generated artifacts, and stores each trial as a consistent snapshot (topology, sizing, placement, routing, and configuration). This trial-level organization enables direct comparison across candidate solutions. For each completed attempt, PANDA runs \textit{PEX} and post-layout simulation, then uses the resulting electrical metrics together with physical check outputs to select the best end-to-end implementation for the target specifications.
\end{sloppypar}

\section{Conclusion and Future Roadmap}
\label{sec:Conclusion}

This paper presented PANDA, an LLM-enhanced performance-driven analog design framework that bridges design intent and layout generation. By integrating topology synthesis, analog sizing, placement, and routing into a coherent flow, PANDA turns high-level requirements into executable circuit and layout actions while preserving cross-stage consistency.

\begin{sloppypar}
Future work will focus on strengthening the learning capability and portability of the framework, including more powerful planning models, tighter coupling between performance estimation and physical optimization, and cleaner service interfaces for back-end tools. These improvements will further advance PANDA toward a more scalable and general analog design automation framework.
\end{sloppypar}

\clearpage
\bibliographystyle{ACM-Reference-Format}
\bibliography{ref/analog}


\end{document}